\documentclass[aps,prb,twocolumn,superscriptaddress]{revtex4}
\pdfoutput=1
\usepackage{graphicx}
\usepackage{rotating}
\usepackage{float}
\usepackage[colorlinks,citecolor=blue,urlcolor=blue, linkcolor=blue]{hyperref}
\usepackage{epstopdf}   % accept eps files for pdflatex compiler
\graphicspath{{figures//}}

\begin{document}   
  
\title{Electronic effects in high-energy radiation damage in iron}
\author{E. Zarkadoula}
\affiliation{Queen Mary University of London, Mile End Road, London, E1 4NS, UK}
\affiliation{South East Physics network}

\author{S. L. Daraszewicz}
\affiliation{London Centre for Nanotechnology, Department of Physics and Astronomy, University College London, Gower Street, London, WC1E 6BT, UK} 

\author{D. M. Duffy}
\affiliation{London Centre for Nanotechnology, Department of Physics and Astronomy, University College London, Gower Street, London, WC1E 6BT, UK}

\author{M. Seaton}
\affiliation{Computational Science and Engineering Department, CCLRC Daresbury Laboratory, Keckwick Lane, Daresbury,
Warrington WA44AD, Cheshire, UK} 

\author{I. T. Todorov}
\affiliation{Computational Science and Engineering Department, CCLRC Daresbury Laboratory, Keckwick Lane, Daresbury,
Warrington WA44AD, Cheshire, UK}

\author{K. Nordlund}
\affiliation{University of Helsinki, P.O. Box 43, FIN-00014 Helsinki, Finland}

\author{M. T. Dove}
\affiliation{Queen Mary University of London, Mile End Road, London, E1 4NS, UK}

\author{K. Trachenko}
\affiliation{Queen Mary University of London, Mile End Road, London, E1 4NS, UK}
\affiliation{South East Physics network}

\begin{abstract}
Electronic effects have been shown to be important in high--energy radiation damage processes where high electronic temperature is expected, yet their effects are not currently understood. Here, we perform molecular dynamics simulations of high-energy collision cascades in $\alpha$-iron using the coupled two-temperature molecular dynamics (2T-MD) model that incorporates both effects of electronic stopping and electron-phonon interaction. We subsequently compare it with the model employing the electronic stopping only, and find several interesting novel insights. The 2T-MD results in both decreased damage production in the thermal spike and faster relaxation of the damage at short times. Notably, the 2T-MD model gives a similar amount of the final damage at longer times, which we interpret to be the result of two competing effects: smaller amount of short-time damage and shorter time available for damage recovery.
\end{abstract}

%\pacs{61.43.Fs, 64.70.Pf, 61.20.Lc}

\maketitle

\section{Introduction}
Structural damage induced by ions carrying energies lower than 100 keV is mainly due to ballistic processes \cite{Ave98}. On the other hand, at high--energy, most of the damage is believed to be due to electronic effects \cite{Kan01b}. Molecular dynamics (MD) simulations have been used to describe radiation damage effects induced when an energetic particle interacts with the matter creating a collision cascade \cite{Rub96,Vic06,Dud08,JAPreview09}. When fast moving atoms interact with the matter, they lose part of their energy due to their interaction with the electrons. The importance of this interaction in the dynamics of a cascade was first mentioned by Flynn and Averback \cite{flynn} and the challenge has been to develop models to include the effects of electronic stopping and the electron--phonon (e-p) interactions in MD simulations. These models include proposals by Caro and Victoria \cite{carovic}, Finnis et al.\cite{finnis}, Ivanov and Zhigilei \cite{zhigi} and Duffy and Rutherford \cite{duffy0}.

Although electronic stopping has been commonly taken into account in
cascade simulations
\cite{kapinos1,kapinos2,zhong,gao,duven1,duven2,duffy1,ismail,phillips,race1,race2,Nor94,Sha94b,Nor98,San09,Bjo09a,Nak10},
there are no systematic studies that include a dynamic,
location-dependent description of how the e-p coupling
affects the atom dynamics in collision cascades. Examining this issue
is especially important in high--energy cascades, where the electronic
excitations matter most. In such events, the high--energy ions lose a
significant amount of their energy due to the inelastic electronic
scattering \cite{duffy0} and high electronic temperatures are
expected. In order to approach high energy events in a more realistic way it is essential to study the effects of the interaction of the atoms with the electrons.

In this paper we study these effects in high--energy cascades in bcc-Fe, a base material for ferritic--martensitic bcc steels, that are the main candidate materials for structural and plasma facing components of future fusion reactors \cite{malerba0,dudder}. We investigate the effect of the e-ph coupling in high energy cascades in bcc--iron by comparing cascades where the energy loss due to electronic stopping has been included in the simulations with cascades where both the electronic stopping and the e-p interaction, as well as the energy feed-back from the electronic to atomic system are included. We are referring to the first set of simulations  
as ``friction cascades'' and to the second set of cascades that implement the full 2T-MD as ``2T-MD cascades''. We see decreased damage production in the thermal spike and faster relaxation of the damage at short times for the 2T-MD cascades. At longer times the 2T-MD model gives a similar amount of final damage, which we interpret to be the combination of two competing effects: smaller amount of short-time damage and shorter time available for damage recovery.  

\section{Methods}
\subsection{The model}

The Duffy and Rutherford 2T-MD model \cite{duffy0,duffy1} is implemented in DLPOLY code \cite{dl1,dl2} version 4.04. It represents the heat exchange between the ionic and electronic subsystems. Inelastic electronic scattering and e-p coupling result in energy loss by the atomic system, which is deposited in the electronic system, where it diffuses and re-deposits energy to the lattice. Thus the electronic system acts as a means for energy transport and storage \cite{duffy4}. The effect of inelastic scattering by the electrons is introduced via a friction term in the equation of motion. The equation of motion has the form of a Langevin equation:

\begin{equation}
m \frac{\partial {\bf v}_{i}}{\partial t} = {\bf F}_{i} (t) - \gamma_{i} {\bf v}_{i} + {\bf \widetilde{F}}(t)
 \label{motion_eq}
\end{equation} 

where m is the mass of atom $i$ and {\bf v$_{i}$} is its velocity. ${\bf F}_{i}(t)$ is the force due to the surrounding atoms of $i$ at time $t$, $\gamma_{i}$ is the friction coefficient and ${\bf \widetilde{F}}(t)$ is a random stochastic force term that is determined by the local temperature of the electronic system (electronic temperature) $T_{e}$. The evolution of the electronic temperature is described by the heat diffusion equation given below (see equation \ref{diffusion}).
 
The friction term is a sum of two parts: a term that accounts for the effect of electron stopping ($\gamma_{s}$) and is applied for velocities of atoms larger than a cut--off value $v_{c}$, and a term that accounts for the e-p interaction ($\gamma_{p}$).

\begin{equation}
\gamma_{i} =  \gamma_{s} + \gamma_{p}~~~~~~\mathrm{for}~~~~{v_{i} > v_{c}}~~~~~
 \label{friction1}
 \end{equation} 

\begin{equation}
\gamma_{i}=\gamma_{p}~~~~~~{\mathrm{for}~~~~{v_{i} \leq v_{c}}}~~~~~
 \label{friction2}
\end{equation}  

The cut--off velocity $v_{c}$ corresponds to energy $E_c$, which in metals is often taken as approximately double the system's cohesion energy \cite{coh2} in order to differentiate ballistically moving atoms (with energy in excess of cohesion energy) from those oscillating. In insulators, it has been shown that the band gap governs the electronic energy losses during the radiation damage process \cite{emilio1,emilio2}.

The magnitude of the stochastic force ${\bf \widetilde{F}}(t)$ is related to the friction coefficient by the fluctuation–-dissipation theorem and the energy exchange drives the atomic system to the temperature of the electronic subsystem \cite{duffy0}. We assume that atoms gain energy only from the e–-p interactions and not from electronic stopping and the stochastic force is proportional only to the e-p interaction friction coefficient $\gamma_{p}$.

\begin{equation}
\langle {\bf \widetilde{F}}(t) \rangle = 0
 \label{fluct1}
 \end{equation}
 
\begin{equation} 
\langle {\bf \widetilde{F}}(t') \rangle . \langle {\bf \widetilde{F}}(t) \rangle = 2 k_{B} T_{e} \gamma_{p}\delta (t' - t).
 \label{fluct2}
 \end{equation}

The MD simulation is coupled to a continuum model for the electronic temperature, which evolves using a heat diffusion equation: 

\begin{equation}
Ce \frac{\partial T_e}{\partial t} = \nabla(\kappa_e \nabla T_e)- g_p (T_e - T_\alpha) + g_s T'_\alpha,
 \label{diffusion}
\end{equation}

where the second and third terms on the right-hand side of the equation represent energy exchange with the lattice via e-p interactions and electronic stopping, respectively. The second term represents energy exchange with the atomic system energy due
to the difference between the atomic system temperature $T_\alpha$ and the electronic system temperature $T_e$. The third term is a source term that describes the energy lost by the atomic system due to electronic stopping. $C_e$ and $\kappa_e$ are electronic specific heat capacity and thermal conductivity, respectively. The atomic temperature $T_\alpha$ is calculated from the average kinetic energy of the atoms in a coarse--grained cell. $T'_\alpha$ has also dimensions of temperature and it is calculated from the average kinetic energy of the subset of atoms with energy greater than twice the cohesive energy of the system \cite{duffy4}. $g_p$ and $g_s$ are the e-p and electronic stopping coupling constants respectively.

The energy loss $\Delta U_{i}$ of an atom i with velocity $v_{i}$ at each timestep with value $\Delta t$ due to a friction force $F_{i}$ is 

\begin{equation}
\Delta U_{i}={\bf F}_{i} {\bf v}_{i} \Delta t = \gamma_{i} v_{i}^2 \Delta t
 \label{U_i}
\end{equation}

In a coarse-grained cell $J$ with constant electronic temperature, the total energy loss will be 

\begin{equation}
\Delta U_{l}= \Delta t \sum\limits_{i \in J} \gamma_{\mathrm{i}} v_{i}^2= \Delta t \sum\limits_{i \in J} \gamma_{\mathrm{p}} v_{i}^2 + \Delta t \sum\limits_{i' \in J} \gamma_{\mathrm{s}} v_{i}^{\prime 2} 
 \label{U_l}
\end{equation}   

where the second sum is over the atoms that have velocities larger than the cut--off velocity that corresponds to double the cohesive energy of the system. The energy gain of the electronic system at each timestep is 

\begin{equation}
\Delta U_{eg}= g_{\mathrm{p}} T_\alpha \Delta V \Delta t + g_{\mathrm{s}} T'_\alpha \Delta V \Delta t .
 \label{U_eg}
\end{equation}  

Equating $\Delta U_{l}$ and $\Delta U_{eg}$ gives  
\begin{equation}
\sum\limits_{i \in J} \gamma_{\mathrm{p}} v_{i}^2 =g_{\mathrm{p}} T_\alpha \Delta V
 \label{equil1}
\end{equation}

\begin{equation}
\sum\limits_{i' \in J} \gamma_{\mathrm{s}} v_{i}^{\prime 2} =g_{\mathrm{s}} T'_\alpha \Delta V
 \label{equil2}
\end{equation}

so $T_\alpha$ and $T'_\alpha$ are defined as 

\begin{equation}
\frac{3}{2} k_{B} T_\alpha = \frac{1}{N} \sum\limits_{i \in J} m v_{i}^2
 \label{T_alpha}
\end{equation} 

\begin{equation}
\frac{3}{2} k_{B} T'_\alpha = \frac{1}{N'} \sum\limits_{i \in J} m v_{i}^{\prime 2}
 \label{T'_alpha}
\end{equation}

and the coupling constants $g_{\mathrm{sp}}$ and $g_{\mathrm{s}}$ as

\begin{equation}
g_{\mathrm{p}} = \frac{3Nk_{B} \gamma_{p}}{ \Delta V m}
 \label{gammap} 
\end{equation} 

\begin{equation}
g_{\mathrm{s}} = \frac{3N'k_{B} \gamma_{s}}{ \Delta V m} 
 \label{gammas}
\end{equation} 

where $N$ is the number of atoms in a coarse-grained cell $J$ with volume $\Delta V$, $k_{\mathrm{B}}$ the Boltzmann constant and $N'$ the number of atoms with velocities larger than $v_{c}$ in the cell $J$.

As described in \cite{duffy0}, the electronic stopping power is proportional to the ion velocity, 

\begin{equation}
\frac{\mathrm{d}E}{\mathrm{d}x} = \mathrm{\lambda} E^{1/2}
\label{propor1} 
\end{equation}

\begin{equation}
m \frac{\mathrm{d}u}{dt} = \mathrm{\lambda} \left(\frac{m}{2}\right)^{1/2} u 
\label{propor2} 
\end{equation}

and the constant of proportionality $\mathrm{\lambda}$ is determined from the Lindhard and Scharff model \cite{lindhard}. Equation \ref{propor2}, from eq. \ref{motion_eq} gives:

\begin{equation}
\gamma_{\mathrm{s}} = \mathrm{\lambda} \left(\frac{m}{2}\right)^{1/2} u 
\label{gamma} 
\end{equation}

and the corresponding relaxation time for electronic stopping is

\begin{equation}
\tau_{\mathrm{s}} = \frac{m}{\gamma_{\mathrm{s}}} = \frac{(2m)^{1/2}} {\mathrm{\lambda}}
\label{tau_s} 
\end{equation} 

The timescale for energy loss due to e-p interactions is 

\begin{equation}
\tau_{\mathrm{p}} = \frac{m}{\gamma_{\mathrm{p}}} 
\label{tau_p} 
\end{equation} 

or from Eq. \ref{gammap} 
\begin{equation}
\tau_{\mathrm{p}} = \frac{3 n k_{B}} {g_{\mathrm{p}}}
\label{tau_p2} 
\end{equation} 

with $n$ being the number of atoms per unit volume.

The heat diffusion equation is solved using a finite difference (FD) method. Energy lost by the atoms, due to the friction term, is input into the local FD cell at each MD timestep. The electronic temperature simulation cell is extended beyond the atomistic simulation cell, and each coarse--grained cell of the electronic grid has length of about 3 \AA. For the electronic grid we are using Robin boundary conditions. We are using a variable timestep for the solution of the FD model, which depends on the electronic temperature gradient and is typically smaller than the MD timestep. The e-p coupling is modelled by a source/sink term in the heat diffusion equation that depends on the difference between the local electronic and lattice temperatures and the e-p coupling constant.  An equivalent amount of energy is removed/added locally to the MD cell by a Langevin thermostat via ${\bf \widetilde{F}}(t)$ that depends on $\mathrm{T}_e$. 

\subsection{Simulations} 

We assume that the e-p coupling process ($g_{p}$) is not initiated until 0.3 ps as the lattice temperature is ill-defined before this. Until this time of the simulation only the electron stopping mechanism is active, while there is a time frame when both the electronic stopping and e-p interaction mechanisms are active. This approximate value was computed by looking at the convergence of kinetic and potential energies (i.e. thermalisation) in the friction cascades. The $C_e(T_e)$ parameterisation was obtained through \emph{ab~initio} calculations as described in \cite{zhigi2}. The heat capacity  given for a range of electron temperatures can be found in \cite{epc}. The temperature dependence of electronic thermal conductivity was assumed to be $\kappa_e(T_e) \sim \frac{C_e(T_e)}{C_e(300~K)}$. In fact we would expect the electronic thermal conductivity to  decrease as the lattice temperature increases but the simple model of the 1/$T_i$ dependence  overestimates this effect as the ionic temperatures can be locally very high. Also the $\tau = 1/T_i$ dependence is neglected in the electron-phonon coupling (g) and therefore for consistency it is neglected in the expression for $\kappa$ (assuming e-p coupling and thermal conduction are linked). Reduced electronic thermal conductivity would contribute two effects: quenching (~1 ps) and annealing at (~1-100 ps) which would potentially decrease the resultant point defect number. However, this does not impact the general conclusion \cite{zhigi,huttner}. We further assume no ionic temperature dependence in $\kappa_e(T_e)$ and a constant value of $g$ \cite{zhigi2}, due to the large uncertainty.

The electronic stopping friction term $\gamma_{s}$ corresponds to value of 1 $\mathrm{ps}^{-1}$ and the cut--off velocity is set to 54 \AA/ps as described in \cite{eza}. 
In this work, in addition to the previously implemented electronic stopping energy loss mechanism, the exchange of the energy between the atomic and the electronic systems is included. The friction coefficient $\gamma_{p}$ due to e-p interactions corresponds to coupling parameter value of \mbox{$g_{\mathrm{p}} = 5.4822 \times 10^{18}$ W m $^{-3}$ K$^{-1}$}   \cite{epc} and is set equal to 1.56 $\mathrm{ps}^{-1}$. 
 A value of $\kappa_{\mathrm{e}} = 80.2$ W m$^{-1}$ K$^{-1}$ for the thermal conductivity at room temperature \cite{li} is used.

We are simulating cascades of 100 keV and 200 keV Fe primary knock--on atoms (PKA) in bcc--Fe in systems that consist of 30, 50 and 100 million atoms. The atoms contained in the boundary of the MD box, in a layer of about 10 \AA\ thickness are connected to a thermostat at 300 K. A variable timestep with a maximum value of \mbox{$1.28 \times 10^{-3}$ ps} is used to describe the atomic motion throughout the cascade development and relaxation. We simulate 12 directions of the PKA on up to 65,000 parallel processors of the HECToR National Supercomputing Service \cite{hector}.

For $\alpha$-Fe, we have used an embedded-atom potential  \cite{mendel}, optimized for better reproduction of several important properties of $\alpha$-Fe, including the energetics of point defects and their clusters ('M07' from \cite{malerba}). At distances shorter than 1 \AA\, interatomic potentials were joined to short-range repulsive ZBL potentials \cite{zbl}. The joining was calibrated against the threshold displacement energies \cite{malerba}. The resulting thresholds were found to be in as good agreement with experiments as the best previous potentials \cite{malerba,Nor05c}.

\section{Results and Discussion}

\begin{figure}[t!]
\begin{flushleft}
\includegraphics[width=\columnwidth]{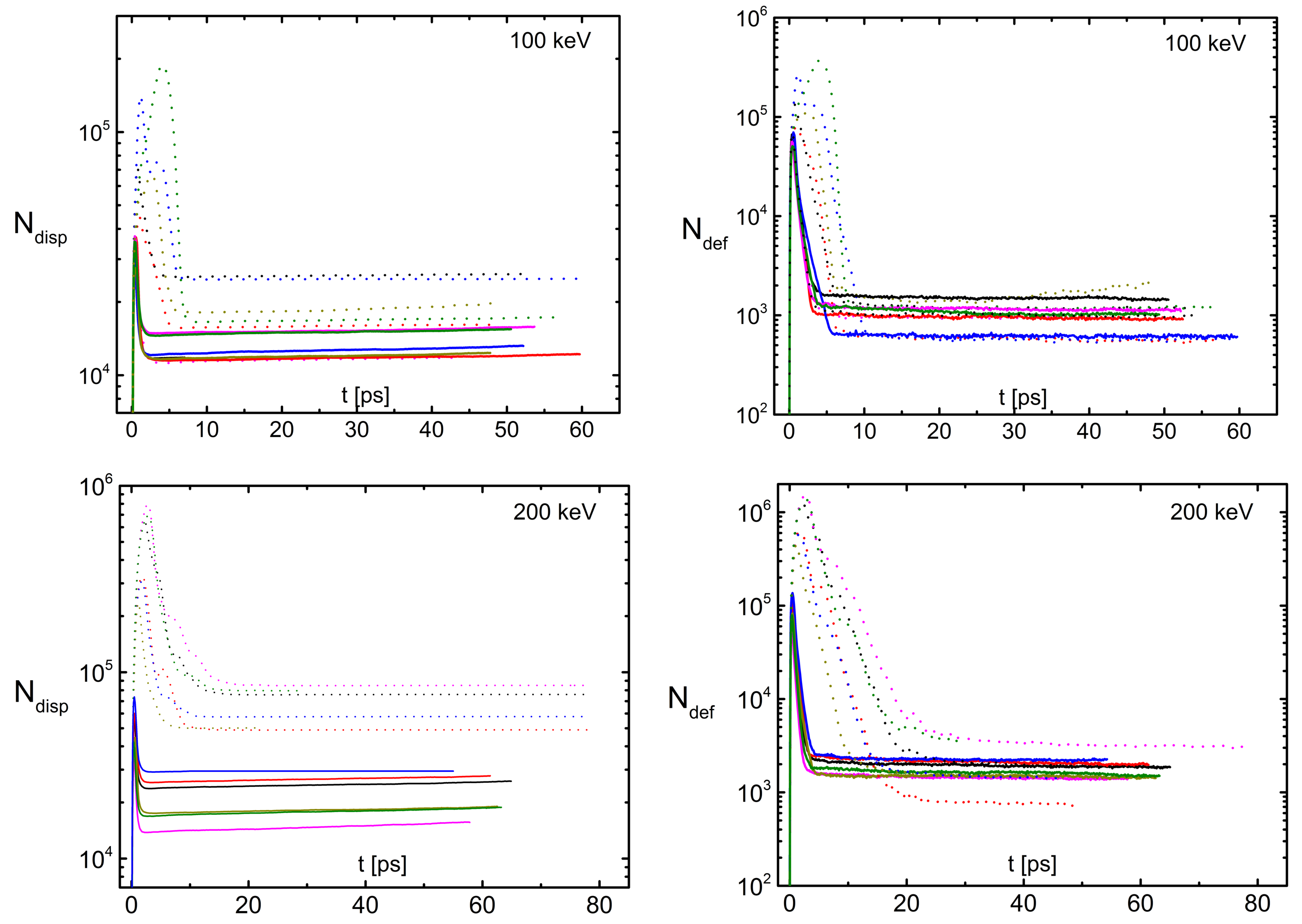}
\end{flushleft}
\caption{$N_{\mathrm{disp}}$ and $N_{\mathrm{def}}$ (sum of interstitials and vacancies) from 100 keV (top) and 200 keV (bottom) knock-on atoms, for different PKA directions. Dotted lines represent the friction cascades, while the solid lines are for the 2T-MD cascades. We see more damage production at the peak for the 2T-MD cascades, and less recombination at the end of the simulation time. Both models result in a similar amount of the final damage at longer times.}
\label{fig1}
\end{figure}

To describe the damage creation and annihilation we first introduce four terms: $N_{\mathrm{disp}}$, $N_\mathrm{def}$, $\tau_{\mathrm{disp}}$ and $\tau_{\mathrm{def}}$. $N_{\mathrm{disp}}$ accounts for the total replacements introduced in the system, i.e. is the number of the atoms that have moved more than a cut--off distance of 0.75 \AA\ from their initial positions. To account for the atoms that recombine to crystalline positions, $N_\mathrm{def}$ is introduced. $N_\mathrm{def}$ reflects the recovery of structural damage as it corresponds to the sum of interstitials and vacancies in the system. An atom is considered an interstitial if it is closer to an occupied lattice side than a cut--off distance $d = 0.75 $ \AA\ and a vacancy is a crystalline position, for which no atom exists closer to this position than the same distance $d$ \cite{kos73} (sphere criterion \cite{Nor97f}). $N_{\mathrm{disp}}$ and $N_\mathrm{def}$ are shown in fig. \ref{fig1} for 100 and \mbox{200 keV} cascades simulated in different knock-on directions.  Specifically, with $N_{\mathrm{disp}}^p$ we refer to the peak of displaced atoms and with $N_\mathrm{def}^p$ to the peak of the defect atoms, often referred to as thermal spike \cite{Bri54,spike,Sam07d}, and $N_{\mathrm{disp}}^l$ and $N_\mathrm{def}^l$ correspond to the number of displaced and defect atoms in long simulation times (the flat lines in fig. \ref{fig1}). $\tau_{\mathrm{disp}}$ is the time during which the elastic recombination of displaced atoms takes place and corresponds to the width of $N_{\mathrm{disp}}^p$ ($N_{\mathrm{disp}}^p$ includes elastic deformation \cite{eza}). $\tau_{\mathrm{def}}$ is the relaxation time during which the dynamic annealing of the defects takes place and corresponds to the width of $N_{\mathrm{def}}^p$.

Corresponding numbers for $N_{\mathrm{disp}}^p$, $N_{\mathrm{def}}^p$, $N_{\mathrm{disp}}^l$ and $N_{\mathrm{def}}^l$ for both simulated energies and for both the friction  and the 2T-MD  models are given in \mbox{Table \ref{tab1}}.

As seen in Fig.\ref{fig1} and Table \ref{tab1}, there is a significant difference in $N_{\mathrm{disp}}^p$ and $N_{\mathrm{def}}^p$ as well as in $\tau_{\mathrm{disp}}$ and $\tau_{\mathrm{def}}$ for both models and simulated energies. First, both $N_{\mathrm{disp}}^p$ and $N_{\mathrm{def}}^p$ are smaller for the 2T-MD cascades as compared to the friction cascades. Second, $\tau_{\mathrm{disp}}$ and $\tau_{\mathrm{def}}$ are shorter for the 2T-MD cascades, corresponding to about 3 ps and 5 ps for $N_{\mathrm{disp}}^p$ and $N_{\mathrm{def}}^p$ respectively, for both simulated energies. These differences are due to faster quenching of the thermal spike in the 2T-MD model that includes the e-p coupling and the additional energy transfer channel. In effect, the e-p coupling removes energy from the thermal spike and electronic thermal conductivity transports it from the simulation cell. This additional energy loss mechanism in the 2T-MD model is also responsible for the smaller amount of unrecombined damage at long times, $N_{\mathrm{disp}}^l$, as is seen in Figure \ref{fig1}. 

\begin{figure}[t!]
\begin{flushleft}
\includegraphics[width=\columnwidth]{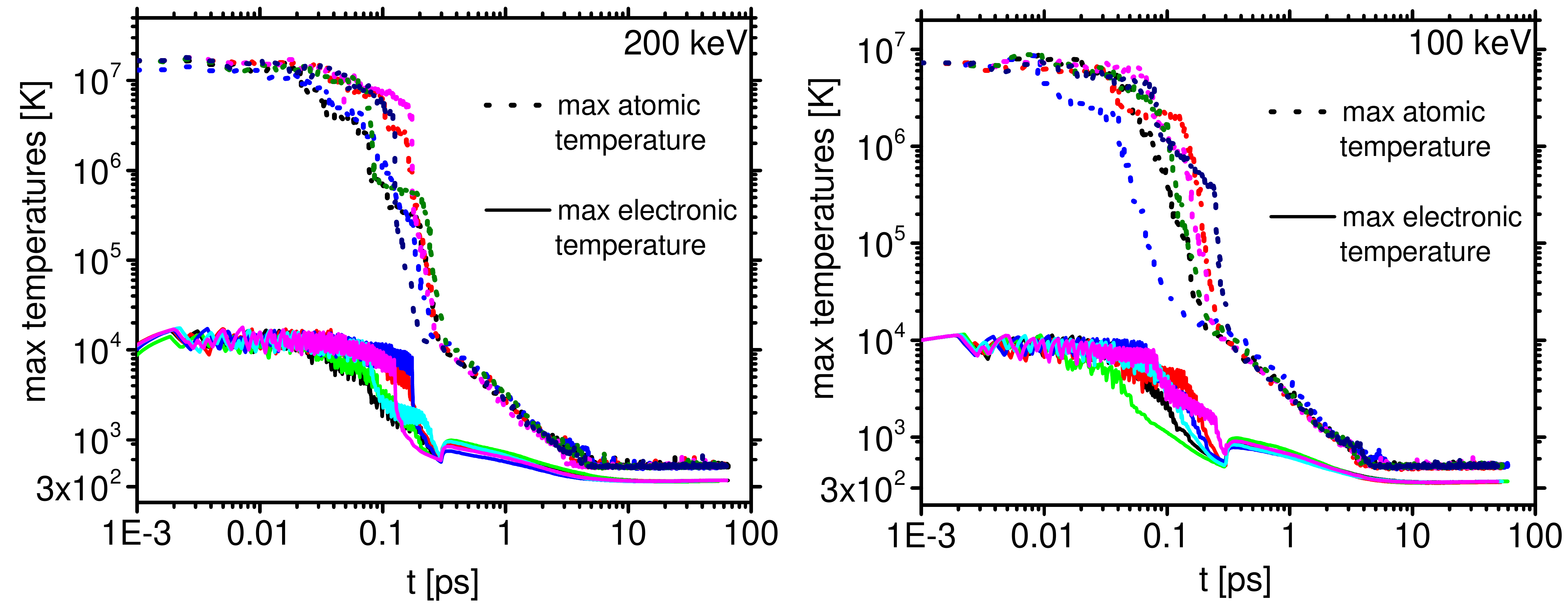}
\end{flushleft}
\caption{Maximum electronic and atomic temperatures for 200 keV (left) and 100 keV (right) 2T-MD cascade simulations, for six events. The ill-defined lattice temperature reaches past $10^7$ K initially. After 0.3 ps, which is the thermalisation time, electronic energy is feedback to the lattice and the ionic temperature starts dropping below $10^4$ K. At around 6 ps the electron-ion temperatures are equilibrated.}
\label{fig2}
\end{figure}

\begin{figure}[t!]
\begin{flushleft}
\includegraphics[width=\columnwidth]{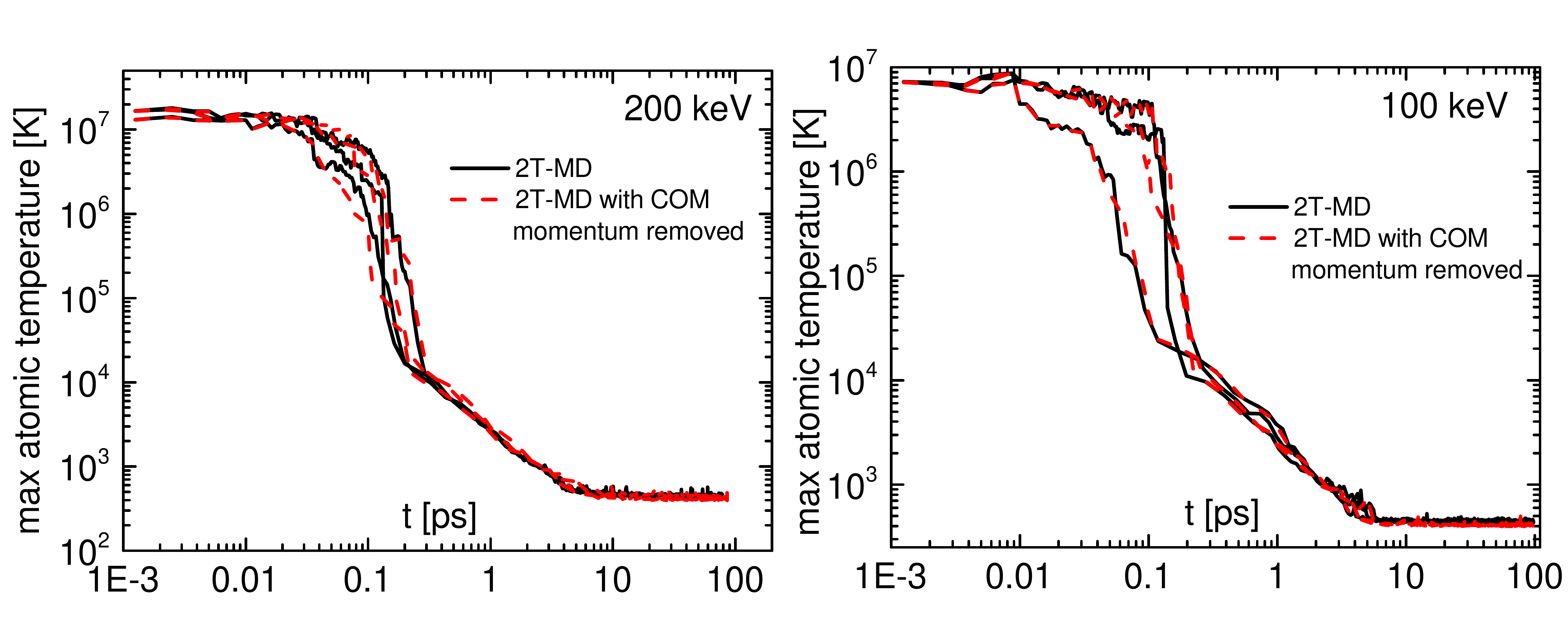}
\end{flushleft}
\caption{Maximum atomic temperatures for 200 keV (left) and 100 keV (right) 2T-MD cascade simulations, for six events. Here we compare simulations where the local COM momentum is subtracted (red dashed lines)  for the local atomic temperature calculation with the original runs, where COM motion is not removed (black solid lines). The runs where local COM momentum is removed do not show a significant difference in comparison to the original runs.} 
\label{fig3}
\end{figure}

\begin{figure}[t!]
\begin{flushleft}
\includegraphics[width=\columnwidth]{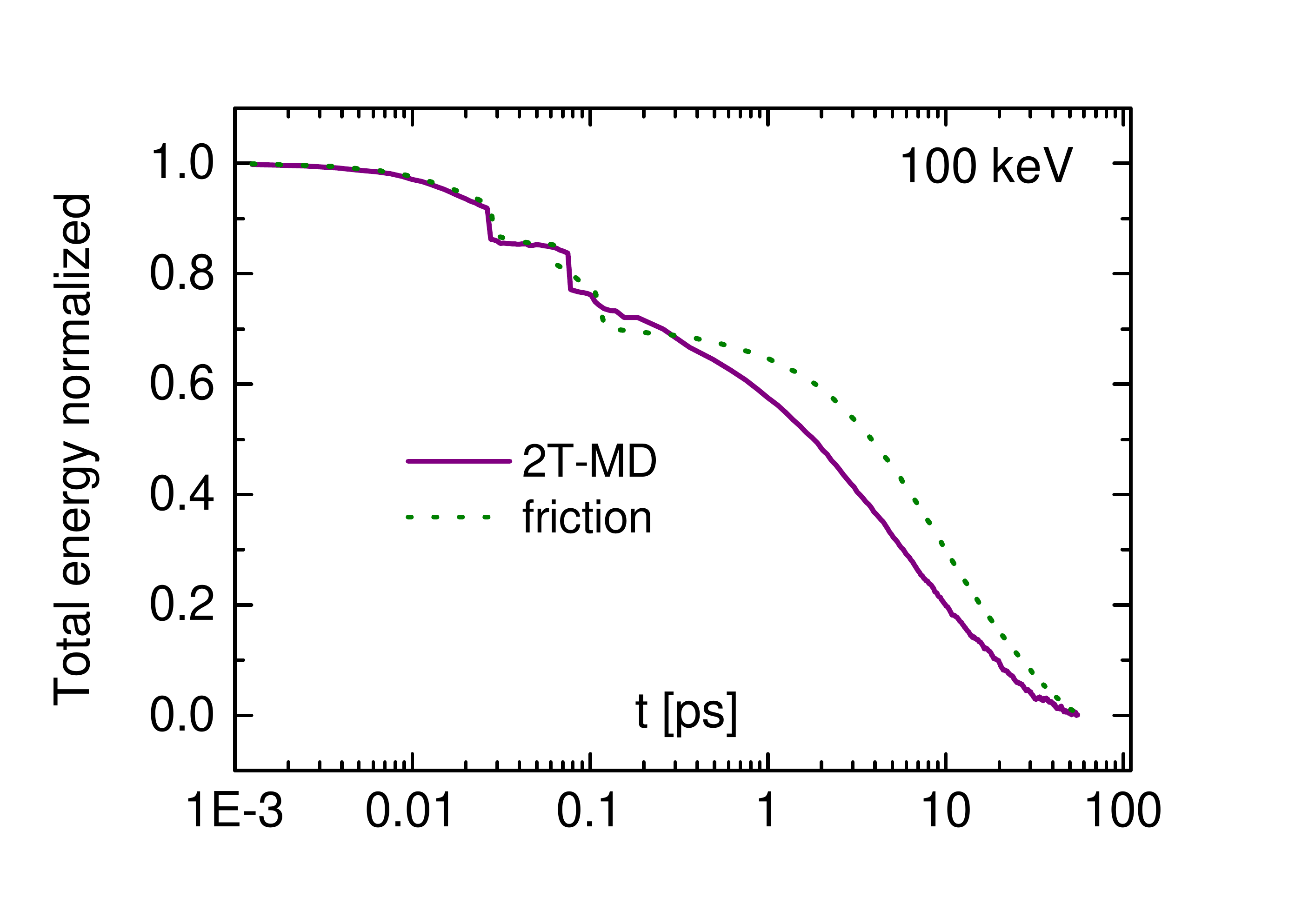}
\end{flushleft}
\caption{Total energy of the system for a representative 100 keV friction cascade (dotted line) and a representative 100 keV 2T-MD cascade (solid line) for the same direction of the PKA. The energy is normalized to 1.}
\label{fig4}
\end{figure}

An interesting insight comes from the examination of $N_{\mathrm{def}}^l$ which quantifies the final amount of damage in the structure and ultimately governs the radiation response of the system. We observe that $N_{\mathrm{def}}^l$ is similar in both 2T-MD and friction models (see Figure \ref{fig1} and Table \ref{tab1}). This effect can be understood on the basis of two competing mechanisms. On one hand, faster energy transfer to the electrons reduces the short--time displaced atoms production, $N_{\mathrm{disp}}^p$. On the other hand, faster energy transfer also reduces the time of the thermal spike in Figure \ref{fig1}, the time that is available for most efficient and fast recombination in the highly mobile and disordered state. As a result, $N_{\mathrm{def}}^l$ in the 2T-MD cascades are similar to $N_{\mathrm{def}}^l$ in the friction cascade where the initial amount of the damage, $N_{\mathrm{def}}^p$ is larger but relaxation time is longer. Figure \ref{fig2} shows the maximum electronic and atomic temperatures for 100 keV and 200 keV 2T-MD cascades, where for all simulations the atomic temperature is higher than the electronic, meaning that the electronic system acts as a heat sink. This is in agreement with lower energy (10 keV) 2T-MD cascades in iron  \cite{duffy0}. The heat transfer relaxation time as read from these plots is about 6 ps. In our simulations we have not subtracted the centre of mass (COM) momentum from each ionic grid cell. We have repeated a number of runs where we removed local COM momentum for calculation of the local ionic temperature and it did not show a significant difference in comparison to the original runs. Comparison of the maximum ionic temperature for the original simulations and the simulations where the COM momentum is removed is shown in Figure \ref{fig3}. Here we see that the maximum ionic temperature for the two methods of calculating the ionic temperature almost coincide. In Figure \ref{fig4} we see the total energy of the system for a 100 keV friction cascade and for a 100 keV 2T-MD cascade for the same direction of the PKA. The energy is normalized to 1. The energy loss is about the same for the two models until 0.3 ps, when the e-p coupling mechanism is activated. For an event of 100 keV, energy of about 30 keV is lost to the electronic system.

\begin{figure}[t]
\begin{center}
\includegraphics[width=\columnwidth]{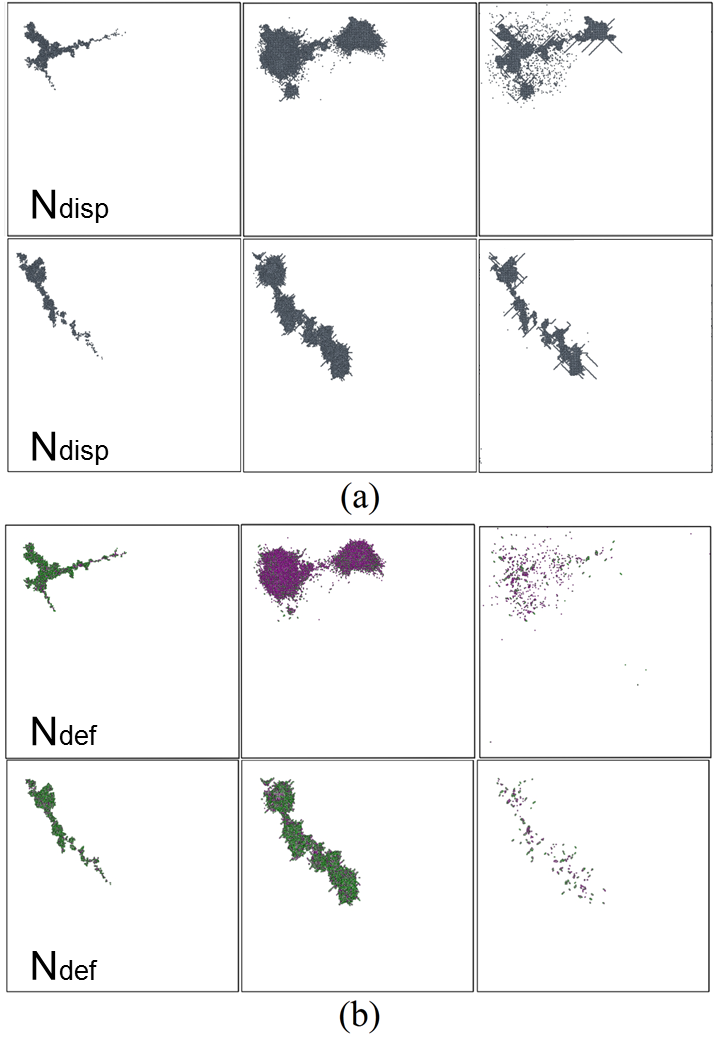}
\end{center}
\caption{Two representative 100 keV cascades for the same PKA direction for the friction mechanism and the full 2T-MD. The PKA moves from the top left to the bottom right corner. (a) Displaced atoms for a friction cascade (top) and for a 2T-MD cascade(bottom). (b) Corresponding defect atoms for the friction (top) and the 2T-MD cascade (bottom). The vacancies (interstitials) are shown in purple (green). The simulation box length is 700 \AA. Friction cascade size and 2T-MD cascade size is 300 \AA\ and 500 \AA, respectively. The snapshots are at 0.1, 2.5 ps and 48 ps for the friction cascade and at 0.1, 0.4 and 53 ps for the 2T-MD cascade. We used Atomeye software \cite{atomeye} to visualize cascade evolution.}
\label{fig5}
\end{figure}

\begin{figure}[thb]
\tiny
\begin{center}
\includegraphics[width=\columnwidth]{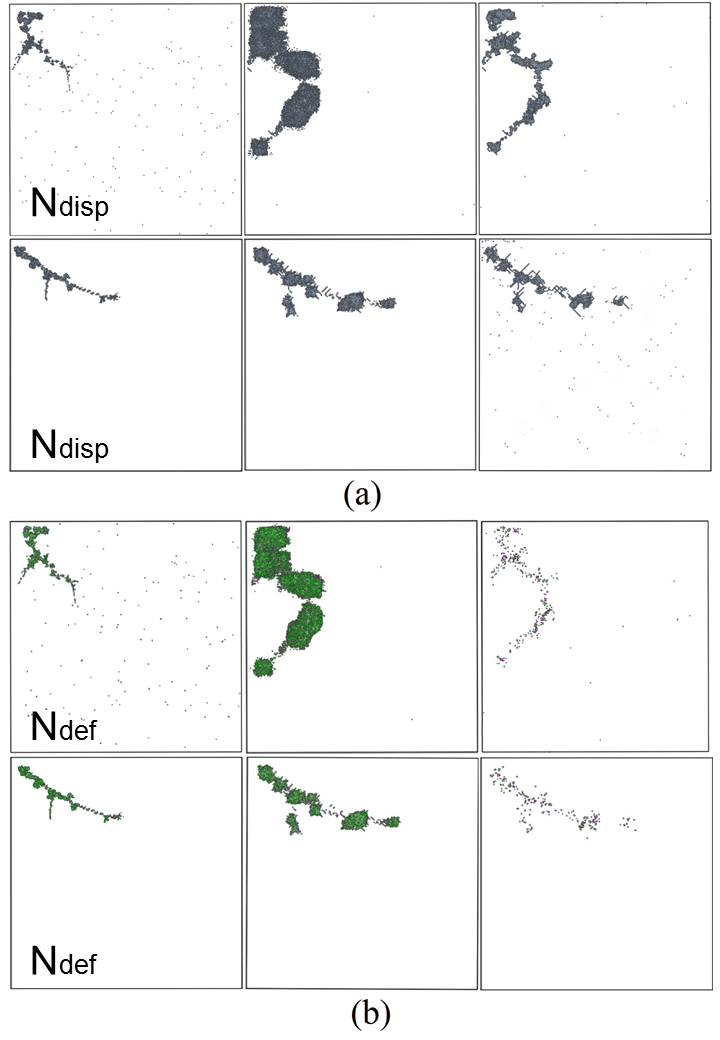}
\end{center}
\caption{Two representative 200 keV cascades for the same PKA direction for the friction mechanism and the full 2T-MD. The PKA moves from the top left to the bottom right corner. (a) Displaced atoms for a friction cascade (top) and for a 2T-MD cascade(bottom). (b) Corresponding defect atoms for the friction (top) and the 2T-MD cascade (bottom). The vacancies (interstitials) are shown in purple (green). The simulation box length is 1100 \AA\ and both friction and 2T-MD cascade size is about 800 \AA. The snapshots are at 0.1, 1.2 ps and 22 ps for the friction cascade and at 0.1, 0.5 and 63 ps for the 2T-MD cascade}
\label{fig6}
\end{figure}

 \begin{table*}[thb]
\setlength{\tabcolsep}{0.2cm}
\begin{tabular}{lcccccc}
&  \multicolumn{2}{c}{\bf PEAK} & \multicolumn{2}{c}{\bf END} \\
   \hline\hline
 {PKA energy} & {\bf $N_{\mathrm{disp}}$}  &{\bf $N_{\mathrm{def}}$} & {\bf $N_{\mathrm{disp}}$}  &{\bf $N_{\mathrm{def}}$} & $\tau_{\mathrm{disp}}$ [ps]& $\tau_{\mathrm{def}}$ [ps]\\
  \hline
100 keV - friction & 89,000 (26,000)  & 146,000 (47,000)  & 19,000 (2,000) & 1,100 (200) & 7 & 10 \\
100 keV - 2T-MD    & 33,000 (2,000)   & 61,000 (3,000)    & 13,000 (700)   & 1,000 (100) & 2 & 5\\
200 keV - friction & 503,000 (98,000) & 982,000 (193,000) & 66,000 (6,000) & 2,000 (400) & 10 & 20\\
200 keV - 2T-MD    & 52,000 (6,000)   &  97,000 (11,000)  & 23,000 (2,000) & 1,700 (100) & 2 & 5\\
\hline\hline
 \end{tabular}
\caption{$N_{\mathrm{disp}}$ and $N_{\mathrm{def}}$, calculated using the sphere criterion, at the peak of the damage (1-2 ps) and at the end of the simulation. Standard error of the mean is shown in the brackets calculated over six events. $\tau_{\mathrm{disp}}$ and $\tau_{\mathrm{def}}$ are read-off from Fig. \ref{fig1}.}
\label{tab1}
\end{table*}
 
\begin{table*}[tb]
\begin{ruledtabular}
\setlength{\tabcolsep}{0.2cm} 
\begin{tabular}{p{2.5cm} p{1.8cm} p{1.5cm} p{1.5cm} p{1.5cm} p{1.8cm} p{1.5cm} p{1.4cm}}
PKA energy & $N_{\rm FP}$ & Number of isolated vacancies & Number of isolated SIAs & Number of vacancy clusters & Number of SIA clusters & Largest vacancy cluster & Largest SIA~cluster \\
  \hline
  100 keV friction    & 550  (200)  & 15 (2)    & 58 (9)    & 26(4)      & 3 (1)   & 18 & 11 \\
  100 keV 2T-MD         & 500  (100)  & 16 (2)    & 68 (6)    & 38(5)      & 6 (1)   & 21 & 12 \\ 
  200 keV friction    & 1000 (400)  & 19 (2)    & 55 (8)    & 75 (6)     & 67 (4)  & 56 & 89 \\
  200 keV 2T-MD         & 850  (100)  & 32 (3)    & 126(9)    & 64 (5)     & 13 (11) & 12 & 8 \\
\end{tabular} 
\end{ruledtabular}
\caption{The number of Frenkel pairs ($N_{\rm FP}$),calculated using the sphere criterion, and defect distribution statistics for 100 keV and 200 keV friction and 2T-MD cascade simulations in $\alpha$-iron. Standard error of the mean is shown in the brackets calculated over six events. The largest clusters that we found in each set of six simulations are presented in the last two columns and are determined by net defect count  which is the difference between the number of self--interstitial atoms  (SIAs) and the number of vacancies.}
\label{tab2}
\end{table*}

In Figures \ref{fig5} and \ref{fig6} we show representative 100 keV and 200 keV cascades for the same PKA direction showing the effects discussed above. Figure \ref{fig5} shows three different time--frames of the relaxation of two representative cascades in a 30 million atoms system, for a 100 keV Fe PKA. Fig. \ref{fig5}(a) shows the displaced atoms for a friction cascade (top) and for a 2T-MD cascade (bottom). Fig. \ref{fig5}(b) shows the defects for the same cascades. The middle frames demonstrate the difference in displaced (about 70,000 for friction and 40,000 for 2T-MD cascade) and defect atoms (115,000 for friction and 70,000 for 2T-MD cascade) for the two mechanisms. The peak for the 2T-MD cascade is at shorter time, 0.4 ps, than for the friction cascade, 2.5 ps demonstrating faster relaxation. $N_{\mathrm{disp}}^l$ corresponds to 20,000 and 15,000 for friction and 2T-MD cascades respectively, as shown in \ref{fig1}. $N_{\mathrm{def}}^l$ is 2,000 for the friction and 2,000 for the 2T-MD cascade. 

Smaller $N_{\mathrm{disp}}^p$, $N_{\mathrm{def}}^p$ and $N_{\mathrm{disp}}^l$ and shorter $\tau_{\mathrm{disp}}$ and $\tau_{\mathrm{def}}$ for 2T-MD cascades are also demonstrated in Fig. \ref{fig6}, where the snapshots of two typical 200 keV collision cascades at three different stages of development are shown. Displaced atoms are shown in fig. \ref{fig6}(a) for a friction cascade (top) and  a 2T-MD cascade (bottom), and defects atoms are shown in fig. \ref{fig6}(b). $N_{\mathrm{def}}^l$ for both mechanisms is 1,500 atoms. For both the 100 keV and the 200 keV cascades shown in fig. \ref{fig5} and \ref{fig6} respectively, we can see that the two models result in different shape of the cascade.

The discussion on the large--scale analysis above focused on the comparison of the dynamics of the two models. Defect analysis results at the local level are summarized in \mbox{Table \ref{tab2}}, where statistics for the defect clusters for the friction and 2T-MD cascades of this paper are given. As discussed above, the difference in the number of Frenkel Pairs (FP) between the two models is small. Similar statistics of defect analysis were obtained for the 100 keV friction and 2T-MD cascades. As shown in the table, we observe statistically significant differences in the defect arrangement in clusters for the 200 keV cascades. 
In particular, the number is isolated vacancies and interstitials is about 
two times higher for the 2T-MD results than for the friction model.
The number of SIA clusters is much smaller for the 2T-MD model at 200 keV. 
This shows that the differences in 2T-MD model and friction
model cascade dynamics (see above) can have significant
effects on damage clustering.

\section{Conclusion}

  Previous works on cascades in Fe have shown that the fraction
of damage in clusters depends both on the interatomic
potential and the way electron-phonon coupling and 
electronic stopping is included in the cascades \cite{Mal09a,Bjo09a}. 
The fraction of damage in large clusters, in turn, may
have a major effect on the long-time scale evolution 
of damage, and is hence a crucial issue for developing
predictive radiation-damage modelling \cite{Bjo11}.
The results in the current work show further that 
treating the electron-phonon coupling in a local way
affects the fraction of damage in clusters. 
Taken together, these results show that assessing accurately the
reliability of primary radiation damage simulations in metals
requires consideration of both the interatomic potentials
and a local model for the electron-phonon coupling. Furthermore, current work on tungsten shows that the results of the e-ph coupling are not generic, which supports that these effects should be investigated for each material individually rather than making extrapolations of results in one material to another, even for cases like iron and tungsten that they both have the bcc structure. The results of the work presented here show that realistic approach of high energy events by treatment of the electronic effects locally is essential, as these effects can significantly affect the arrangement of the defects in clusters, and therefore they can affect the long term performance of the material.

\end{document}